\documentclass{elsart}
\usepackage{graphicx}

\usepackage{amssymb}

\def\etal{et al.~}
\def\eg{e.g.,~}
\def\ie{i.e.,~}
\def\kms{~{\rm km~s^{-1}}}
\def\cm3{~{\rm cm^{-3}}}

\def\lsim{\mathrel{  
        \raise0.3ex\hbox{$<$}\kern-0.75em{\lower0.65ex\hbox{$\sim$}}}}
\def\gsim{\mathrel{
        \raise0.3ex\hbox{$>$}\kern-0.75em{\lower0.65ex\hbox{$\sim$}}}}

\begin{document}

\begin{frontmatter}



\title{Self-Similar Evolution of Cosmic-Ray-Modified Quasi-Parallel Plane Shocks}


\author[label1]{Hyesung Kang}
\ead{kang@uju.es.pusan.ac.kr}
\author[label2]{T. W. Jones}
\ead{twj@astro.umn.edu}
\ead[url]{www.astro.umn.edu/$\sim$twj}

\address[label1]{Pusan National University, Pusan 609-735, Korea}
\address[label2]{University of Minnesota, Minneapolis, MN 55455, USA}

\begin{abstract}
Using an improved version of the previously introduced CRASH 
(Cosmic Ray Acceleration SHock) code,
we have calculated the time evolution of cosmic-ray (CR) modified 
quasi-parallel plane shocks for Bohm-like diffusion,
including self-consistent models of Alfv\'en wave drift and
dissipation, along with ``thermal leakage injection'' of CRs.
The new simulations follow evolution of the CR distribution to
much higher energies than our previous study, providing a better
examination of evolutionary and asymptotic behaviors. 
The postshock CR pressure becomes constant
after quick initial adjustment,
since the evolution of the CR partial pressure expressed
in terms of a momentum similarity variable is self-similar. 
The shock precursor, which
scales as the diffusion length of the highest energy CRs, subsequently
broadens approximately linearly with time, independent
of diffusion model, so long as CRs continue to be
accelerated to ever-higher energies. 
This means the nonlinear shock structure can
be described approximately in terms of the similarity
variable, $x/(u_s t)$, where $u_s$ is the shock speed 
once the postshock pressure reaches an approximate time asymptotic state. 
As before, the shock Mach number
is the key parameter determining 
the evolution and the CR acceleration efficiency, although
finite Alfv\'en wave drift and wave energy dissipation in the
shock precursor reduce the effective velocity change experienced
by CRs, so reduce acceleration efficiency noticeably,
thus, providing a second important parameter at low and moderate Mach
numbers.
For low Mach numbers ($M_0\lsim 5$) the CR acceleration efficiency 
depends on the thermal leakage injection rate,
the Alfv\'enic Mach number, and any preexisting CR population.
However, these dependences become weak for high shock Mach numbers 
of $M_0>30$.  To evaluate CR acceleration efficiencies in the
simulated shocks we present for a wide range of shock parameters
a ``CR energy ratio'', $\Phi(M_0)$, comparing
the time asymptotic volume-integrated energy in CRs to the time-integrated
kinetic energy flux through the shock. This ratio asymptotes
to roughly $0.5$ for sufficiently strong shocks. The postshock CR pressure
is also approximately 1/2 the momentum flux through the shock
for very high Mach numbers.

\end{abstract}

\end{frontmatter}

\section{Introduction}

Astrophysical plasmas, from the interplanetary gas inside the heliosphere
to the galaxy intracluster medium (ICM), are magnetized and turbulent
and contain nonthermal particles in addition to gas thermal particles. 
So, understanding complex interactions among these different components
is critical to the study of many astrophysical problems.
In collisionless shocks entropy is generated via collective 
electromagnetic viscosities, \ie interactions of charged particles 
with turbulent fields \cite{maldru01}.
Some suprathermal particles of the shock heated gas can leak upstream,
their streaming motions against the background fluid
exciting MHD Alfv\'en waves upstream of the shock \cite{bell78,lucek00}.
Then those particles can be further accelerated to very high energies
through multiple shock crossings resulting from resonant scatterings with the 
self-excited Alfv\'en waves in the 
flows converging  across the shock \cite{dru83,blaeic87,maldru01}.

Detailed nonlinear treatments of diffusive shock acceleration (DSA) 
account for incoming thermal particles injected
into the CR population (\eg \cite{ell85,malvol98,gies00}) 
as a consequence of incomplete 
thermalization by collisionless dissipation processes. 
Those particles, while relatively few in number, can subsequently accumulate
a major fraction of the shock kinetic energy as their
individual energies increase \cite{ebj95,kjg02}.
Such predictions are supported by a variety of observations
including direct measurements of particle spectra at interplanetary 
shocks, 
nonthermal $\gamma$-ray, X-ray and radio emissions of supernova remnant shocks and
also possibly the ICM of some X-ray clusters (\eg \cite{blaeic87,aharon04,rott97}).
CR acceleration may be universal to astrophysical shocks in diffuse,
ionized media on all scales.

Unlike an ideal gasdynamic shock, downstream states of a CR modified shock
cannot be determined in a straightforward way by simple jump conditions 
across the shock front.
This is because the shock transition depends on the CR pressure
distribution upstream of the dissipative subshock. The
particle acceleration takes place on diffusion time
and length scales ($t_d(p)=\kappa(p)/u_s^2$ and $l_d(p)=\kappa(p)/u_s$,
where $\kappa(p)$ is diffusion coefficient and $u_s$ is the shock speed),
which are much larger than the shock dissipation scales.
Unless or until some boundary condition limits the maximum
CR momentum, this structure will continue to evolve along with
the CR distribution.
Thus the evolution of a CR shock with a finite age and size should properly
be followed by time-dependent numerical simulations.
In addition, complex interplay among CRs, resonant waves, and the underlying gas flow 
(i.e., thermal leakage injection, self-excited waves, resonant scatterings of
particles by waves, and non-linear feedback to the gas flow) is 
model dependent and not yet understood completely.

In the time dependent kinetic equation approach to numerical study of CR acceleration
at shocks, the diffusion-convection equation for the particle momentum
distribution, $f(p,x,t)$, is solved, along with suitably modified gasdynamic
equations (\eg \cite{kjls01}).
Since accurate solutions to this equation require 
a computational grid spacing smaller than the particle diffusion length,
$l_d(p)$,
and since realistic diffusion coefficients have steep momentum dependence,
a wide range of length scales must be resolved
in order to follow the CR acceleration from the injection momentum
(typically $p_{\rm inj}/m_pc \sim 10^{-2}$) to highly relativistic
momenta ($p/m_pc \gg 1$). This constitutes an extremely challenging
numerical task, which can require rather extensive computational resources,
especially if one allows temporal and spatial evolution of 
diffusion behavior.
To overcome this numerical problem in a generally applicable way
we have built the
CRASH (Cosmic-Ray Acceleration SHock) code 
by implementing Adaptive Mesh Refinement (AMR) techniques and 
subgrid shock tracking methods \cite{kjls01,kj06} 
in order to enhance computational efficiency.
The CRASH code also treats thermal leakage injection self-consistently
by adopting a shock transparency function  for suprathermal
particles in the shock \cite{kjg02}.

We previously applied our CRASH code in a plane-parallel geometry 
to calculate the nonlinear evolution of CR modified shocks in the
absence of significant local Alfv\'en wave heating and
advection \cite{kjg02,kj02,kj05}.
For those models the shock sonic Mach number, $M_0$, largely
controlled the thermal leakage injection rate and the CR acceleration
efficiency in evolving modified planar shocks, 
since $M_0$ determines the relative velocity jump across the shock and 
consequently the degree of shock modification by CRs.
In all but some of the highest Mach number shocks 
the CR injection rate and the postshock CR pressure 
approached time-asymptotic values when a balance was achieved between 
acceleration/injection and diffusion/advection processes. 
This resulted in an approximate ``self-similar'' flow structure,
in the sense that the shock structure broadened approximately
linearly in time, so that the shock structure could be
expressed in terms of the similarity coordinate $u_s t$.
It is likely that all of the models would have reached such asymptotic dynamical
structures eventually, but performance limits
in the version of the code in use at that time prevented us
from extending some of the simulations long enough to confirm that. The 
CR distribution evolved only to $p_{\rm max}/m_pc \sim 10-10^3$. 
Based on the self-similar evolution reported in our previous
work, we calculated the ratio of CR energy 
to inflowing kinetic energy, $\Phi$, (see Eq. [12] below) as a measure of the CR acceleration efficiency
in a time-asymptotic limit.
The CR energy ratio, $\Phi$, increased with the shock Mach number, but 
approached $\approx 0.5$ for large shock Mach numbers, $M_0 > 30$,
and it was relatively independent of other upstream properties or 
variation in the injection parameter.
In those shocks where we observed time asymptotic dynamical
behaviors the postshock CR pressures were $\sim$30 - 60\% of the
ram pressure in the initial shock frame, this ratio increasing with
Mach number. For some of the highest Mach number shocks in that study
CR pressure continued to increase to the end of the simulation, so
the final values could not be measured. 
Finally, the presence of a preexisting, upstream CR population was
seen in those earlier simulations to be equivalent to having
slightly more efficient thermal leakage injection for such strong shocks,
while it could substantially increase the overall CR energy in moderate
strength shocks with $M_0<3$.

In the present paper, we revisit the problem of self-similar evolution of
CR modified shocks with a substantially improved numerical scheme that enables us to
follow the particle acceleration to energies much higher than we
considered before.  
This allows us to measure asymptotic dynamical
properties for all the newly simulated shocks and
to demonstrate that the self-similar evolution of the
CR partial pressure in terms of a momentum similarity variable
leads to the constancy of the postshock CR pressure. 
We also include in the new work the effects of Alfv\`en wave
drift and dissipation in the shocks.
The time asymptotic CR acceleration efficiency is once again
controlled by the shock Mach number but diminished as the ratio of
Alfv\'enic Mach number to sonic Mach number decreases.
The asymptotic shock properties are largely independent of the
magnitude of the spatial diffusion coefficient and also its
subrelativistic momentum dependence.

The basic equations and details of the numerical method are described
in \S 2.
We present simulation results for a wide range of shock parameters 
in \S 3, followed by a summary in \S 4.

\section{Numerical Method}
\subsection{Basic Equations}
The evolution of CR modified shocks depends on a coupling between
the gasdynamics and the diffusive CRs. That coupling takes place by way of
resonant MHD waves, although it is customary to express
the pondermotive wave force and dissipation in the plasma
through the associated CR pressure distribution properties along with
a characteristic wave propagation speed (usually the Alfv\'en speed) (\eg \cite{skill75,ach82}).
Consequently, in our simulations we solve the standard gasdynamic 
equations with CR pressure terms
added in the conservative, Eulerian
formulation for one dimensional plane-parallel geometry. 
The evolution of a modified entropy, $S = P_g/\rho^{\gamma_g - 1}$,
is followed everywhere except across the subshock,
since for strongly shocked flows,
numerical errors in computing the gas pressure from the total
energy can lead to spurious entropy generation with
standard methods, especially in the shock precursor \cite{kjg02}.
 
\begin{equation}
{\partial \rho \over \partial t}  +  {\partial (u \rho) \over \partial x} = 0,
\label{masscon}
\end{equation}
 
\begin{equation}
{\partial (\rho u) \over \partial t}  +  {\partial (\rho u^2 + P_g + P_c) \over
\partial x} = 0,
\label{mocon}
\end{equation}

\begin{equation}
{\partial (\rho e_g) \over \partial t} + {\partial \over \partial x} 
(\rho e_g u + P_g u) = 
-u {{\partial P_c}\over {\partial x}} 
\label{econ}
\end{equation}

\begin{equation}
{\partial S\over \partial t}  +  {\partial\over \partial x} (S u) =
+ {(\gamma_{\rm g} -1)\over \rho^{\gamma_{\rm g} -1} } [W(x,t) - L(x,t)], 
\label{scon}
\end{equation}

where $P_{\rm g}$ and $P_{\rm c}$ are the gas and the CR pressure,
respectively, $e_{\rm g} = {P_{\rm g}}/{[\rho(\gamma_{\rm g}-1)]}+ u^2/2$
is the total energy of the gas per unit mass.
The remaining variables, except for $L$ and $W$ have standard meanings.
The injection energy loss term, $L(x,t)$, accounts for the
energy carried by the suprathermal particles injected into the CR component at
the subshock and is subtracted from the postshock gas 
immediately behind the subshock. 
Gas heating due to Alfv\'en wave dissipation in the upstream region is 
represented by the term
$W(x,t)= - v_A {\partial P_c / \partial x }$, 
where $v_A= B/\sqrt{4\pi \rho}$ is the Alfv\'en speed.
This commonly used dissipation expression derives from a quasi-linear 
model in which Alfv\'en waves are amplified by
streaming CRs and dissipated locally as heat in the precursor region
(\eg \cite{jon93}).

The CR population is evolved by solving the diffusion-convection equation in the form,
\begin{equation}
{\partial g\over \partial t}  + (u+u_w) {\partial g \over \partial x}
= {1\over{3}} {\partial \over \partial x} (u+u_w)( {\partial g\over
\partial y} -4g) + {\partial \over \partial x}  [\kappa(x,y)  
{\partial g \over \partial x}], 
\label{diffcon}
\end{equation}
where $g=p^4f$, with $f(p,x,t)$ the pitch angle averaged CR 
distribution, 
and where $y=\ln(p)$, while $\kappa(x,p)$ is the spatial diffusion coefficient
\cite{skill75}.
For simplicity we always express the particle momentum, $p$ in
units $m_{\rm p}c$ 
and consider only the proton CR component.
The wave speed is set to be $u_w=v_A$ in the upstream region, while we
use $u_w=0$ in the downstream region.
This term reflects the fact that
the scattering by Alfv\'en waves tends to isotropize 
the CR distribution in the wave frame rather than the bulk-flow, gas frame 
\cite{skill75}. Upstream, the waves are expected to be dominated by the
streaming instability, so face upwind. Behind
the shock, various processes, including wave reflection, are
expected to lead to a more nearly isotropic wave field (\eg \cite{achbl86}).

Eqs. (1)-(5) are simultaneously integrated by the CRASH code in 
plane-parallel geometry. The detailed numerical description can be found in Kang 
\etal 2002 \cite{kjg02}.
A key performance feature of the CRASH code is multiple levels of 
refined grids (typically $l_g=8-10$) strategically laid around the subshock 
to resolve the diffusion length scale of the lowest energy particles near 
injection momenta. 
Grid refinement spans a region around the subshock just large 
enough to include comfortably the diffusion scales of dynamically 
important high energy CRs with enough levels to follow freshly injected 
low energy CRs with sufficient resolution to produce converged
evolutionary behaviors.
To accomplish grid refinement effectively it is necessary to locate
the subshock position exactly. Thus, we track the subshock
as a moving, discontinuous jump inside the initial, uniform and
fixed grid \cite{kjls01}.

\subsection{Diffusion Model}

We considered in this study two common choices for diffusion models.
First is the Bohm diffusion model, which represents scattering expected
for a saturated wave spectrum and gives what is generally assumed to be
the minimum diffusion coefficient
as $\kappa_{\rm B} = 1/3 r_{\rm g} \upsilon$,
when the particles scatter within a path of one gyration radius 
(\ie $\lambda_{\rm mfp} \sim r_{\rm g}$). 
This gives
\begin{equation}
\kappa_B(p) = \kappa_{\rm n} {{p^2}\over {(p^2+1)^{1/2}}}.
\label{bohm}
\end{equation}
The coefficient
$\kappa_{\rm n} = m c^3/(3eB) = (3.13\times 10^{22} {\rm cm^2s^{-1}} ) B_{\mu}^{-1}$,
where $B_{\mu}$ is the magnetic field strength in units of microgauss.
There has been much discussion recently about amplification of the
large scale magnetic field within the shock precursor (\eg \cite{lucek00,maldi06,vlad06}).
Since physical models of that evolution are still not well
developed, we will assume for simplicity in the simulations presented here
that the large scale field is constant through the shock structure.

Because of its steep momentum dependence in the nonrelativistic regime,
the Bohm diffusion model requires an extremely fine spatial grid resolution
whenever nonrelativistic CRs are present.
On the other hand the form of $\kappa(p)$ for nonrelativistic
momenta mostly impacts only early evolution of CR-modified
shocks, when CR feedback is dominated by nonrelativistic particles
and thermal leakage injection rates are adjusting rapidly to
changes from initial conditions.
So, to concentrate computational effort more efficiently, we adopted
in some previous works \cite{kjls01,kj06} a ``Bohm-like'' diffusion
coefficient that includes a weaker momentum dependence
for the non-relativistic regime, 
\begin{equation}
\kappa_{BL}(p) = \kappa_{\rm n} p. 
\end{equation}
According to those previous studies, the differences in results
between the two models are minor except during early nonlinear shock
evolution, as expected.
Thanks to the weaker momentum dependence of $\kappa_{BL}$ we can, 
for given computational resources, calculate numerically converged models with smaller $\kappa_n$ 
resulting in the acceleration of CRs to higher momenta.

In order to quench the well-known CR acoustic instability in the precursor of highly modified
CR shocks (\eg \cite{kjr92}), we assume a density dependence for the
diffusion coefficient, $(\rho/\rho_0)^{-1}$, so that
$\kappa(x,p) = \kappa_B (\rho/\rho_0)^{-1}$
or $\kappa(x,p) = \kappa_{BL} (\rho/\rho_0)^{-1}$,
where $\rho_0$ is the upstream gas density.
This density dependence also models enhancement of the
Alfv\'en wave magnetic field amplitude due to flow compression.
We note, also, for clarity that hereafter we use the subscripts '0', '1', and '2' to denote
conditions far upstream of the shock, immediately upstream of the
gas subshock and immediately downstream of the subshock, respectively.

\subsection{Thermal Leakage Model}

In the CRASH code suprathermal particles are injected as CRs 
self-consistently via ``thermal leakage'' through the lowest CR momentum boundary.
The thermal leakage injection model emulates the filtering process by which
suprathermal particles well into the tail of the postshock Maxwellian 
distribution leak upstream across the subshock \cite{malvol98,maldru01}.
This filtering is managed numerically by
adopting a ``transparency function'', $\tau_{\rm esc}(\epsilon_B, \upsilon)$,
that expresses the probability of supra-thermal
particles at a given velocity, $\upsilon$, successfully swimming
upstream across the subshock through the 
postshock MHD waves \cite{gies00,kjg02}.
The one model parameter, $\epsilon_B = B_0/B_{\perp}$, is the ratio of the amplitude
of the postshock wave field interacting with the low energy
particles, $B_{\perp}$, to the general magnetic field, $B_0$,
which is aligned with the shock normal in these simulations.
The transparency function fixes the lowest momentum of the CR component
in our simulations from the condition that $\tau_{\rm esc}>0$ 
(\ie non-zero probability to cross the subshock for CRs) for $p>p_1$, where 
$p_1=(u_2/c) (1+\epsilon_B)/\epsilon_B$ and $u_2$ is the downstream
flow speed in the subshock rest frame.
Initially $p_1$ is determined by the downstream speed of the initial shock, 
but it decreases as the subshock weakens and then it becomes constant after
the CR modified shock structures reach asymptotic states.

Since suprathermal particles have to swim 
against the scattering waves advecting downstream, 
the subshock Mach number, $M_s$,
is one of the key shock characteristics that control the injection fraction. 
Previous simulations showed that injection is less efficient 
for weaker shocks, but becomes independent of $M_0$ for
strong shocks, when the subshock compression asymptotes \cite{kjg02}.
For a given total shock Mach number, $M_0$, on the other hand,
the injection rate is controlled mainly by the parameter $\epsilon_B$ 
In practice we have found that
$\epsilon_B\sim 0.2-0.25$ leads to an injection fraction 
in the range $\sim 10^{-4}-10^{-3}$. This is similar to commonly adopted values in
other models that employ a fixed injection rate
(\eg \cite{berz97,mal97,ama05}). 
Although somewhat higher field turbulence values 
($0.25<\epsilon_B<0.3)$ are suggested 
theoretically for strong shocks \cite{mal98}, these 
evoke a start-up problem in the numerical simulations,
since they lead to very rapid initial injection that cools the
postshock flow too strongly for it to remain numerically stable. 
Once the shock structure becomes nonlinear, however,
those influences moderate greatly, so that, as we have shown
previously, the ultimate CR acceleration behavior
depends only weakly on $\epsilon_B$.
In fact, we have found previously for strong shocks that the time-asymptotic
behaviors are very weakly dependent on $\epsilon_B$ \cite{kjls01}, so
its chosen value will have no influence on our final conclusions.

We directly track the fraction of particles
injected into the CR population as follows: 
\begin{equation}
\xi(t) =  { {\int {\rm dx} \int_{p_1}^{p_2} 4\pi f_{\rm CR}(p,x,t)p^2 {\rm dp}} 
\over {n_0 u_{s,0} t}}, 
\end{equation}
where $f_{\rm CR}$ is the CR distribution function above $p_1$, 
while ${n_0 u_{s,0} t}$ is the number of particles passed through
the shock until the time $t$.
The highest momentum of the CR component, $p_2$, is chosen so that it is
well above $p_{\rm max}$ at the simulation termination time,
where $p_{\rm max}$ is defined in \S 2.4.

\subsection{CR Acceleration Efficiency}

The postshock thermal energy in a gasdynamic shock can, of course, be 
calculated analytically by the Rankine-Hugoniot jump condition. On the
other hand, the CR population and the associated acceleration efficiency
at CR modified shocks should properly be 
obtained through time-dependent integration of the shock
structure from given initial states, since the CR distribution
depends on the shock structure, which 
is not discontinuous and continues to evolve so long as the CR 
population evolves.
In particular CR modified shocks contain a smooth precursor in the upstream
region whose scale height grows in time in proportion to the
diffusion length of energetically dominant particles. 
The total shock compression may, similarly, evolve over a 
significant time period.
The standard expression for the mean acceleration timescale for a 
particle to reach momentum $p$ in the test-particle limit of DSA theory 
is given by \cite{lc83} 
\begin{equation}
\tau_{acc}(p)
= {3\over {u_1-u_2}} ({\kappa_1\over u_1} + {\kappa_2\over u_2}).
\end{equation}
In the test particle limit the shock compression is given by
the Rankin-Hugoniot condition. Then assuming a $\gamma_{\rm g} = 5/3$ gasdynamic
shock and a diffusion coefficient
taking the density dependence indicated at the end of \S 2.2, this leads to
\begin{equation}
\tau_{acc}(p) \approx 8 { M_0^2 \over {M_0^2-1}} {\kappa(p) \over u_s^2}, 
\label{tacc}
\end{equation}
where $u_s$ and $M_0$ are the shock speed and sonic Mach number, respectively.
While this expression should strictly speaking be modified in
highly modified CR shocks, since the shock structure and the
associated CR transport 
are more complex than assumed in Eq. (\ref{tacc}) \cite{blasi06},
we find it empirically to be reasonably consistent with our results
described below. Accordingly, we may expect and confirm below that
the time-dependent evolution of our CR modified shocks will be determined
primarily by the shock Mach number and can be expressed simply in terms of
diffusion length and time scales. 

Within this model the highest momentum expected to be accelerated 
in strong shocks by the time $t$
is set according to Eq. (\ref{tacc}) by the relation $t\approx 8 \kappa(p_{\rm max})/u_s^2$. 
In that case the scale height of the precursor 
or shock transition structure grows linearly with time as 
\begin{equation} 
 l_{\rm shock} \sim {\kappa(p_{\rm max}) \over u_s} \sim {1\over 8}{u_s t}, 
\end{equation} 
independent of the magnitude or the momentum dependence of $\kappa(p)$.
This evolution should continue until some other physics limits
the increase in CR momentum, such as the finite size of the shock system.
Since the CR pressure approaches a time-asymptotic value (see Figs. \ref{fig3}-\ref{fig5} below),
the evolution of the CR-modified shock becomes, under these circumstances,
approximately self-similar, independent of the form of the diffusion 
coefficient, 
while $l_{\rm shock}$ grows linearly with time \cite{kj02,kang03,kj05}.
On the other hand, $\kappa(p_{\rm max}) \approx l_{\rm shock} u_s$,
so for Bohm-like diffusion, $p_{\rm max} \approx l_{\rm shock}(t) u_s/\kappa_n$. 
Thus, at a given time the CR distribution, $g(p)$, extends for
Bohm-like diffusion according to
$p_{\rm max} \approx  u_s^2 t/(8 \kappa_n)  \propto 1/\kappa_n$.

\begin{figure}
\begin{center}
\includegraphics*[height=40pc]{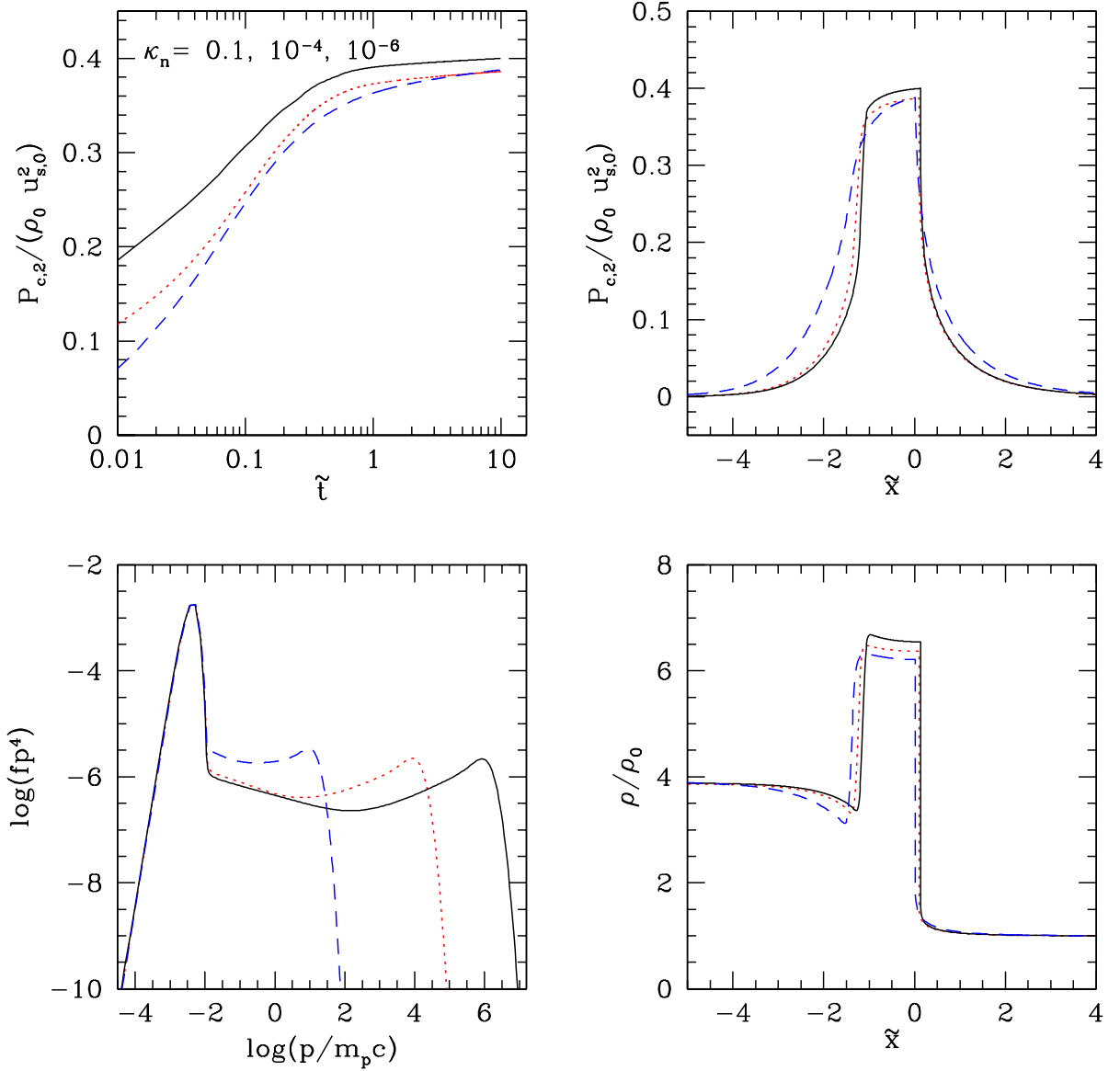}
\end{center}
\caption{
{\it Upper left panel}: Time evolution of the postshock CR pressure for
three shocks with $M_0=20$, $T=10^6$K and $\theta = 0.1$.
Three models with different values of $\tilde \kappa_n=0.1$ (dashed line), 
$10^{-4}$ (dotted line), and $10^{-6}$ (solid line) are shown for
comparison.
{\it Lower left panel}: The CR distribution function at the subshock,
$g(p,x_s)=f(p,x_s)p^4$,
is plotted for the three models at $\tilde t=10$.
Also spatial profiles of the CR pressure and density for
the three shock models are shown in the 
upper right and lower right panels, respectively.
The normalization scales are defined in the text.}
\label{fig1}
\end{figure}

Fig. \ref{fig1} shows a comparison of three models of a $M_0=20$ shock with
$\tilde \kappa_B=\tilde \kappa_n p^2/\sqrt{p^2+1}$ using $\tilde \kappa_n = 0.1$, 
and with $\tilde \kappa_{BL} = \tilde \kappa_n p$
using $\tilde \kappa_n = 10^{-4}~ {\rm and}~10^{-6}$ in units
defined in the following section.
The upper left panel, displaying the evolution of postshock CR pressure,
demonstrates similar time-asymptotic values for all three models. 
At early times the CR pressure evolution depends on
details of the model, including numerical properties
such as spatial and momentum grid resolutions, and the previously
described injection suppression scheme used to prevent start up problems.

The other panels in Fig. \ref{fig1} illustrate shock structure comparisons for
the three models at the end of the simulations. 
The shock structures, $\rho(x)$ and $P_c(x)$, are very similar,
while the CR spectrum extends to different values of $p_{\rm max}$,
inversely proportional to $\tilde \kappa_n$.

The self-similar evolution of CR modified shocks makes it useful
to apply the ratio of CR energy to a fiducial kinetic
energy flux through the shock as a simple measure of acceleration 
efficiency; namely,
\begin{equation}
\Phi(t)=\frac {\int E_c(x,t) {\rm d x}} { 0.5\rho_0 u_{s,0}^3  t }.
\label{crenrat}
\end{equation}
More specifically, this compares the total CR energy within the simulation box to
the kinetic energy in the {\it initial shock frame} that has crossed 
the shock at a given time. As the shock structures approach
time-asymptotic forms the above discussion suggests
that $\Phi(t)$ also may approach
time-asymptotic values. This is confirmed in our
simulations. We see also that the asymptotic $\Phi$ values
depend in our simulations primarily on
shock sonic Mach number and are independent of $\kappa$.

The highest momenta achieved in our simulations are set by
practical limits on computation time controlled by the vast
range of diffusion times and lengths to be modeled.
Still, the asymptotic acceleration efficiency ratio is almost 
independent of the maximum momentum reached.
For the three models shown in Fig. \ref{fig1}, for example, 
$p_{\rm max} \sim 10,~10^4,$ and $10^6$ at $\tilde t = 10$, 
depending on the value of $\kappa_n$, but 
the CR energy ratio approaches similar values of $\Phi\sim 0.4$ 
for all three models.


\section{Simulation Set Up and Model Parameters}

\subsection{Units and Initial Conditions}

The expected evolutionary behaviors described above naturally suggest 
convenient units. For example,
given a suitable velocity scale, $\hat u$, and length scale, $\hat x$, a
time scale $\hat t = \hat x/\hat u$ and diffusion coefficient
scale, $\hat \kappa = \hat x \hat u = \hat u^2 \hat t$ are implied. Alternatively,
one can select the velocity scale along with a convenient scale for the 
diffusion coefficient, $\hat \kappa$, leading to a natural
length scale, $\hat x = \hat \kappa/\hat u$ and a related time scale,
$\hat t =  \hat \kappa/\hat u^2 = \hat x/\hat u$. We will follow the latter
convention in our discussion. 
In addition, given an arbitrary mass unit, $\hat m$, which we take to be
the proton mass, we can similarly normalize mass
density in terms of $\hat \rho = \hat m/\hat x^3$. 
Pressure is then expressed relative to $\hat \rho \hat u^2$. 
For clarity we henceforth indicate
quantities normalized by the above scales using a tilde; for example, $\tilde u$, $\tilde t$,
and $\tilde \kappa_n$. 

We start each simulation with a pure gasdynamic, right-facing shock at rest
in the computational grid. We use the upstream gas speed in this
frame, $u_0$ as our velocity scale, so that the initial, normalized
shock speed is $\tilde u_{s,0} = 1$ with respect to the upstream gas.
The upstream gas is specified with one of two temperature 
values, either $T_0=10^4$K 
or $T_0=10^6$, which represent warm
or hot phase of astrophysical diffuse media, respectively.
In astrophysical environments, for example, photoionized gas of $10^4$ K is
quite common.  Hot and ionized gas of $>10^6$ K is also found in
the hot phase of the ISM \cite{McKee77,Spitzer90}. 
The shock speed and the upstream temperature are related through
the sonic Mach number, $M_0$, by the usual relation
$u_{s,0} = c_{s,0} M_0 = 15\kms(T_0/10^4)^{1/2} M_0 = u_0$,
where $c_{s,0}$ is the sound speed of the upstream gas.
So, by choosing $T_0$ and $M_0$, we set the physical value of the shock
speed, which, in turn, determines the postshock thermal behavior.
For CR distribution properties it is also necessary to define the 
speed of light, $c$, in terms of $ \beta_k = u_0/ c$. 
The normalized upstream gas density is set to unity; \ie $\tilde \rho_0 = 1.$

The preshock pressure is determined by the shock Mach number,
$\tilde P_{\rm g,0}= (1/\gamma_{\rm g})M_0^{-2}$,
where the gas adiabatic index, $\gamma_{\rm g}=5/3$.
The postshock states for the initial shocks are determined by the 
Rankine-Hugoniot shock jump condition.
For models with $T_0=10^4$K, $M_0=$10-80 is considered,
since the gas would not be fully ionized at slower shocks
($u_s<150 \kms$), the postshock gas would often become radiative and
the CR acceleration become inefficient owing to wave dissipation 
from ion-neutral collisions (\eg \cite{ddk96}). 
For models with $T_0=10^6$K $M_0=2-30$ ($300\kms \le u_s \le
4500 \kms$) is considered,
since the CR acceleration should be relatively independent of $T_0$
for shocks with $M_0>30$. 

In order to explore effects of pre-existing CRs, 
we also consider, as we did in our earlier work, models with
$T_0=10^6$ K that include an ambient (upstream) CR population, 
$f(p)\propto p^{-5}$ for $p_1\le p \le p_2$
and set its pressure $P_{c,0}=(0.25-0.3)P_{g,0}$.
For these models, we adopt $\kappa_B=0.1p^2/\sqrt{p^2+1}$,
$p_1=(u_{s,0}/c)(1+\epsilon_B)/\epsilon_B$ and $p_2=10^3$.
For strong shocks the presence of pre-existing CRs is
similar in effect to having a slightly higher injection rate,
so the time asymptotic shock structure and CR acceleration efficiency 
depend 
only weakly on such a pre-existing CR population \cite{kjg02,kang03}. 
On the other hand, for weak shocks, a pre-existing CR pressure comparable
to the upstream gas pressure represents a
significant fraction of the total energy entering the
shock, so pre-existing CRs obviously have far more impact.
In addition, the time asymptotic CR acceleration efficiency in weak shocks depends
sensitively on the injection rate, so increases with increased
shock transparency, controlled through $\epsilon_B$ (see \S 2.3).
Hence, as we found in \cite{kj05}, we expect relatively weak CR shocks ($M_0<5$) to
be substantially altered by the presence of a finite upstream $P_c$.

\subsection{Wave Drift and Heating}

\begin{figure}
\begin{center}
\includegraphics*[height=40pc]{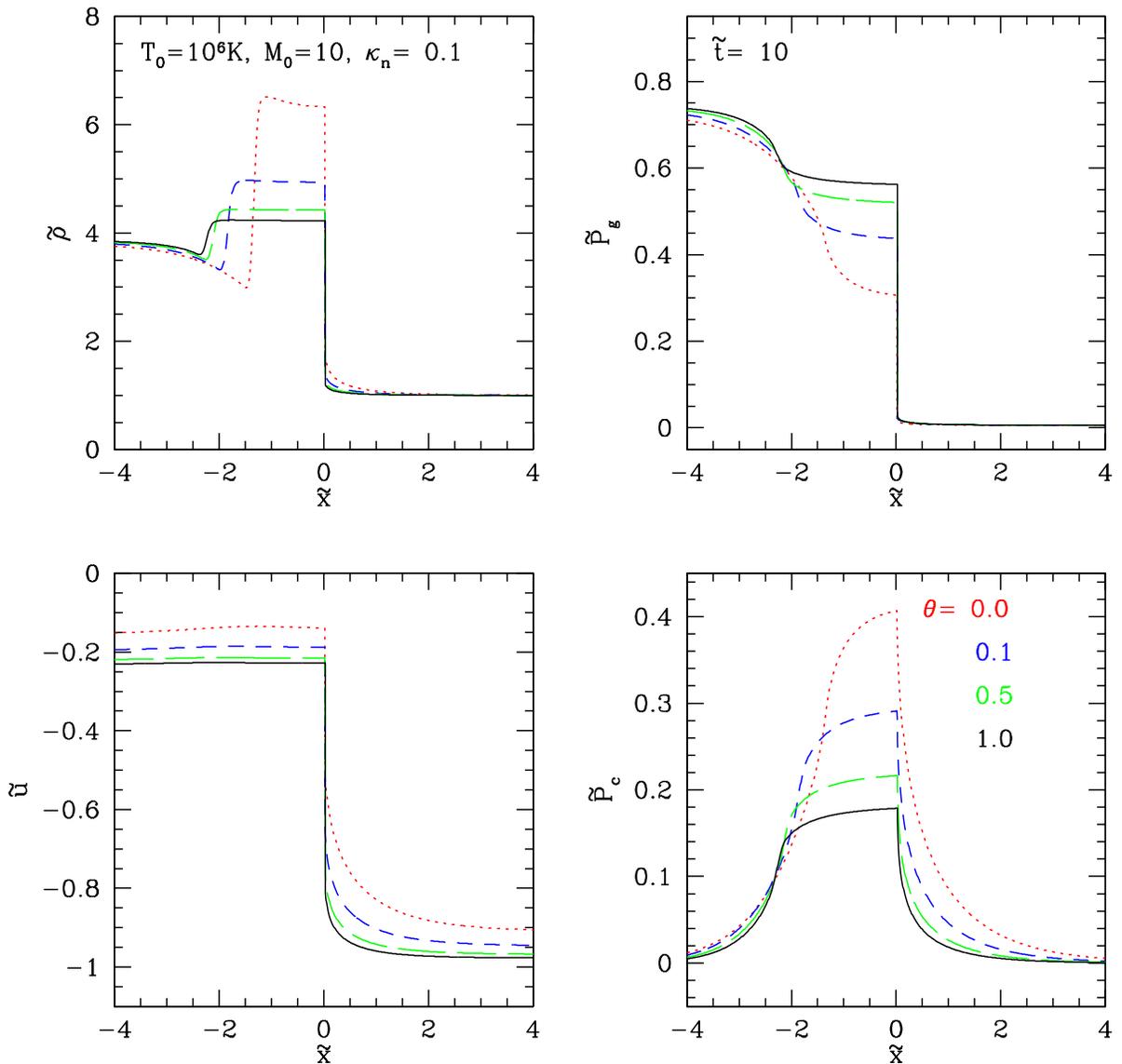}
\end{center}
\caption{Time asymptotic structure of a $M_0 = 10$ CR-modified
shock illustrating the role of Alfv\'en wave
drift and dissipation in determining shock properties.
Four models with $\theta=0$ (dotted line), 0.1 (dashed),
0.5 (long dashed), and 1.0 (solid) are shown. 
As described
in the text, $\theta=E_B/E_{th,0} = (5/9) (M_0/M_A)^2$.}
\label{fig2}
\end{figure}

As shown in earlier works (\eg \cite{jon93,kj06} and references
cited therein),
the CR acceleration becomes less efficient when Alfv\'en wave drift and 
heating terms are included in the simulations.
This behavior comes from two effects previously
mentioned in \S 2.1, both of which derive
from the resonance interaction between CRs and Alfv\'en waves
in the shock precursor. The Alfv\'en waves stimulated
by CR streaming in the precursor will propagate in the upstream
direction, so that the effective advection
speed of the CRs into the subshock is reduced. In addition,
if the energy extracted from CRs to amplify these waves is
locally dissipated, the heating rate in the precursor is
increased with respect to the adiabatic rate, so that gas entering
the subshock is relatively hotter, and the subshock strength is
accordingly reduced. The significance of the effects depends on
the sonic Mach number, $M_0$, relative to 
the shock Alfv\'enic Mach number, $M_A = u_s/v_A$; 
\ie on the ratio of the Alfv\'en speed to the sound speed (see \S 3.4). 
In a parallel shock we can write 
$M_0 /M_A = v_A/c_s = \sqrt{ 2\theta/\left[\gamma_{\rm g} 
(\gamma_{\rm g} - 1)\right]}$, 
where we introduce a convenient ``Alfv\'en parameter'' as follows,
\begin{equation} 
\theta = { {E_{B,0}}\over{E_{th,0}}} = (\gamma_{\rm g} - 1) {P_{g,0}\over P_{B,0} }= 
{(\gamma_{\rm g} - 1) \over \beta_p}.
\end{equation} 
This expresses the relative upstream Alfv\'en and sound speeds in terms
of the magnetic
to thermal energy density ratio. The parameter $\beta_p$ is the usual
``plasma $\beta$'' parameter. For $\gamma_{\rm g} = 5/3$, $\beta_p = 2/(3\theta)$.
and $\theta = (5/9) (M_0/M_A)^2$.
We emphasize for clarity that the present simulations 
are of parallel shocks, so that the direct dynamical role of
the magnetic field has been neglected.
Observed or estimated values are typically $\theta \sim 0.1$ for
intracluster media and $\theta \sim 1$ for the interstellar medium of our
Galaxy (\eg \cite{beck01,car02}).  So we consider $0.1\le \theta \le 1$, and
we will provide comparison to the weak field limit, $\theta = 0$.  
Fig. \ref{fig2} concisely illustrates the importance of Alfv\'en wave
drift and heating effects on a Mach 10 shock. One can see that
the postshock CR pressure, for example, decreases by more than a 
factor two when the sonic and Alfv\'en Mach numbers become
comparable. Most of the model results we show here used
$\theta = 0.1$. For the Mach 10 shocks illustrated in Fig. \ref{fig2}
the associated wave terms have reduced the asymptotic CR pressure
by about 30\% with $\theta = 0.1$ compared to the shock with no such terms included.

\subsection{Grid Resolution and Convergence}

According to our previous studies \cite{kj91}, the spatial grid resolution 
should be much finer than the diffusion length of lowest energy particles
near the injection momenta, \ie $ \Delta x \lsim 0.1 l_d(p_{\rm inj})$. 
Kang \& Jones (2006), however, showed that the spherical, comoving CRASH code 
achieve good numerical convergence, even when $ \Delta x > l_d(p_{\rm inj})$.
This was due to the fact in the execution of that code
the gas subshock remains consistently inside the same comoving grid zone.
In order to gain this benefit for our present simulations,
we have modified our plane-parallel CRASH code so that 
the shock is again forced to remain inside the same refined grid zone 
by regularly redefining the underlying Eulerian grid.
The simulations employ eight levels of refinement with the grid spacing reduced by an integer
factor of two between refinement levels. 
The spatial grid resolution on the coarsest, base grid is $\Delta \tilde x_0 = 2\times 
10^{-3}$, while on the finest, $8^{th}$, grid 
$\Delta \tilde x_8 = 7.8\times 10^{-6}$. 
With $\tilde \kappa_n=10^{-6}$, the diffusion length for injection momenta 
$p_1 \approx 10^{-2}$ becomes $\tilde l_d \approx 10^{-8}$
for models with Bohm-like diffusion, $\kappa_{BL}$.
Although $\Delta \tilde x_8> \tilde l_d(p_{\rm inj})$, 
we confirmed that the new plane-parallel CRASH code also achieves good 
numerical convergence in the simulations presented here. 
This improvement enables us to
extend these simulations to CR momenta several orders of magnitude greater
that those discussed in \cite{kj05}.

When solving the diffusion-convection equation, we used 230 - 280 uniformly spaced 
logarithmic momentum bins in the interval $y=\ln p = [\ln p_1,\ln p_2]$.

\subsection{Results}

\begin{figure}
\begin{center}
\includegraphics*[height=40pc]{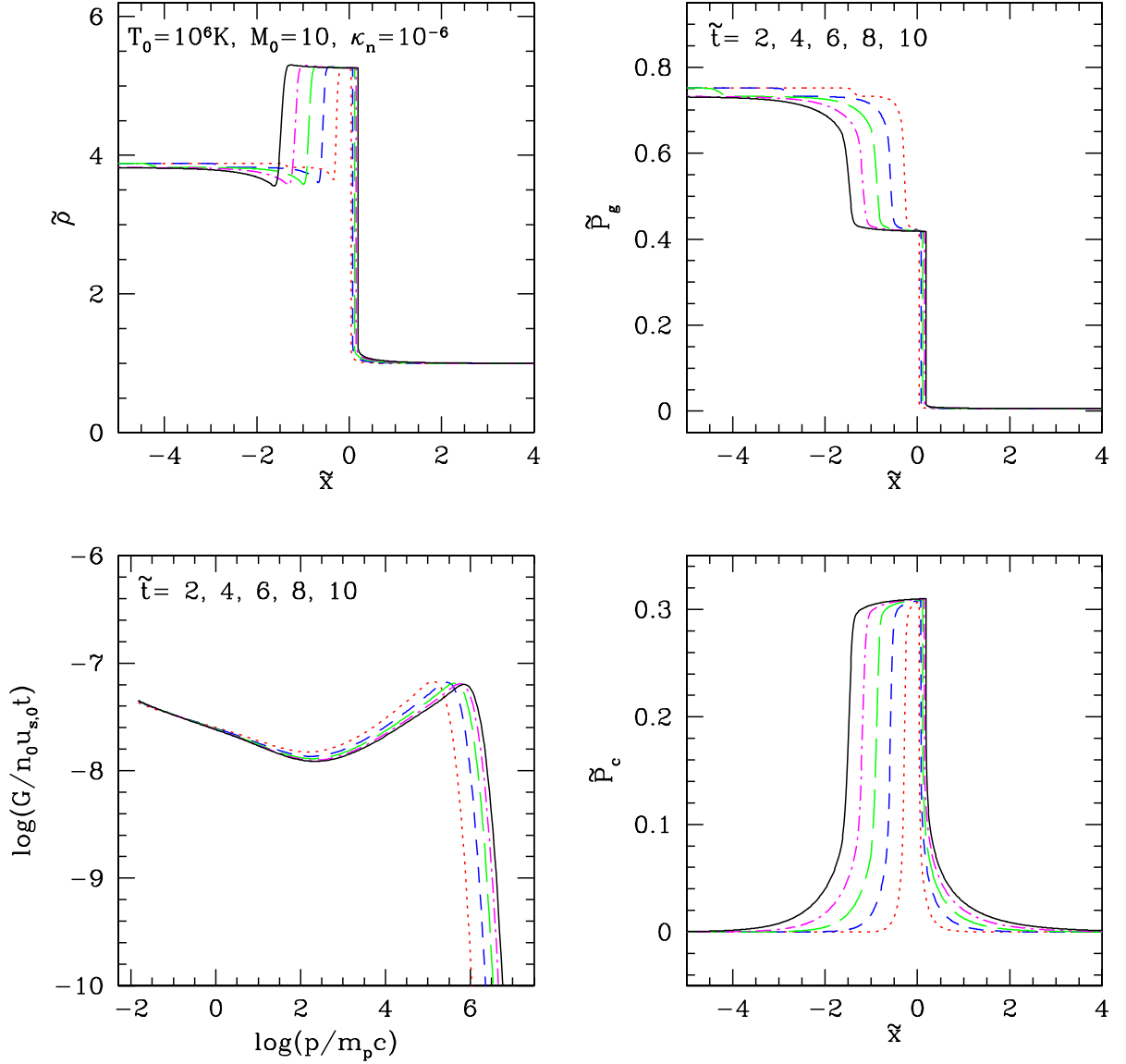}
\end{center}
\caption{
Evolution of a plane-parallel shock with $M_0=10$ and $T=10^6$K
($u_s=1500\kms$) is shown for $\tilde t = 2, 4,..10$.
The initial, $\tilde t = 0$, gas shock is set at rest in the computational grid
by Rankine-Hugoniot shock jump condition.
The injection parameter for thermal leakage injection is 
$\epsilon_B=0.2$ and the diffusion coefficient is $\tilde \kappa_{BL}=10^{-6}p$.
The Alfv\'en parameter, $\theta = 0.1$.
The lower left panel shows the volume integrated CR spectrum,
$G(p) = \int dx f(x,p)p^4 $ in terms of the number of particles 
passed through the shock, $n_0 u_{s,0} t$. 
}
\label{fig3}
\end{figure}

We show in Fig. \ref{fig3} the time evolution of a $M_0=10$ shock with $T_0=10^6$K, 
$\tilde \kappa_{BL}=10^{-6}p$, and $\theta =0.1$.
The wave amplitude parameter in the thermal
leakage model was assumed to be $\epsilon_B=0.2$,
unless stated otherwise.
The lower left panel follows evolution of the volume integrated CR distribution function
relative to the total number of particles (mostly in the thermal population) that have passed through the shock,
\ie $G(p)/(n_0u_{s,0} t)$, where $G(p) = \int {\rm dx} g(x,p)$.
As the CR pressure increases in the precursor in response to thermal leakage 
injection at the subshock and subsequent Fermi acceleration, 
the subshock weakens.
The injection process is self-regulated in such a way
that the injection rate reaches and stays at a nearly stable value after
quick initial adjustment.
Consequently, the postshock CR pressure reaches an approximate time-asymptotic value 
once a balance is established between fresh injection/acceleration and
advection/diffusion of the CR particles away from the shock.

\begin{figure}
\begin{center}
\includegraphics*[height=40pc]{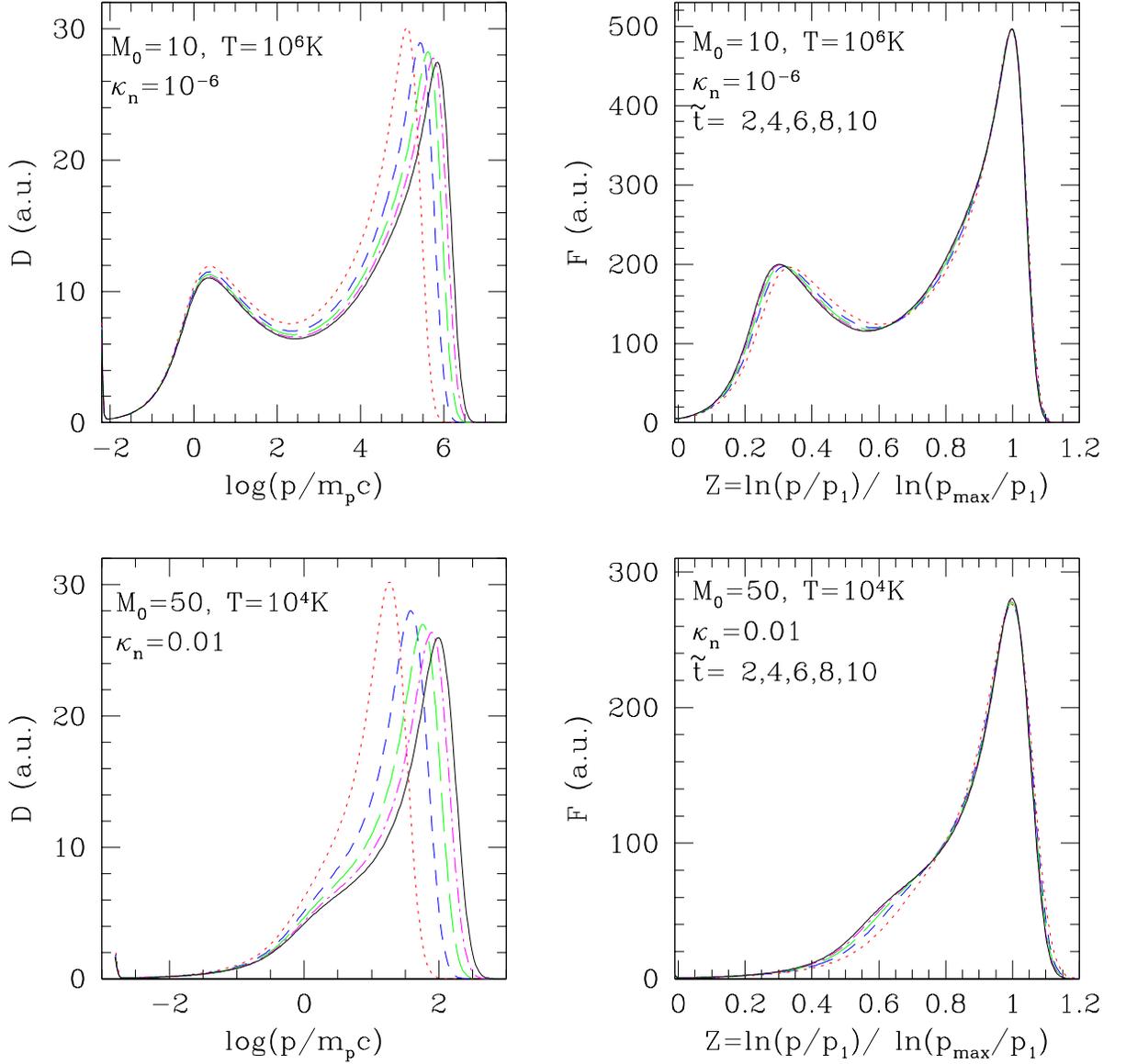}
\end{center}
\caption{
The partial pressure functions $D(p,x_s)$ and $F(Z,x_s)$ (in arbitrary
units) at the subshock are shown at $\tilde t= 2, 4,...10$.
Upper panels are for the model shown in Fig. 3,
while lower panels are for the model with 
$M_0=50$, $T_0=10^4$K, $\tilde \kappa_{B}=0.01p/\sqrt{p^2+1}$
and $\theta=0.1$.
See text for the definitions of the momentum similarity
variable $Z$ as well as $D(p)$ and $F(Z)$.
}
\label{fig4}
\end{figure}

The CR pressure is calculated as 
\begin{equation}
P_c = {{4\pi}\over3} m_p c^2 \int_{p_1}^{p_2} g(p) {p \over \sqrt{p^2+1}} d\ln p,
\end{equation}
so we define $D(p) \equiv g(p)p/\sqrt{p^2+1}$ as a `partial pressure function'. 
The upper left panel of Fig. \ref{fig4} shows the evolution of $D(p,x_s)$ 
at the subshock for the model shown in Fig. \ref{fig3}.
Since $D(p)$ stretches self-similarly in momentum space, we define a new
momentum similarity variable as
\begin{equation}
Z \equiv {\ln(p/p_1) \over \ln[p_{\rm max}(t)/p_1]}.
\end{equation}
The lowest momentum $p_1$ becomes constant after the subshock structure
becomes steady. 
The momentum at which numerical values of $D(p)$ peaks is chosen 
as $p_{\rm max}(t)$ and similar to what is estimated approximately
by applying the test-particle theory in \S 2.4. 
For the model shown in Fig. 3,  
$p_1 \approx 0.01$ and $p_{\rm max} \approx 7.27\times 10^4~ \tilde t$. 

We then define another `partial pressure function',
\begin{equation}
F(Z) \equiv g(Z) { p \over \sqrt{p^2+1}}  \ln[p_{\rm max}(t)/p_1]
=D(Z) \ln[p_{\rm max}(t)/p_1].
\end{equation}
Its time evolution is shown as a function of $Z$ at the upper right panel 
of Fig. \ref{fig4}.
The plot demonstrates that the evolution of $F(Z)$ 
becomes self-similar for $\tilde t \ge 2$.
Since $P_c \propto \int_{p_1} D(p) d\ln p \propto \int _0 F(Z) dZ$,
the areas under the curves of $D(p)$ or $F(Z)$ in Fig. 4 
represent the CR pressure at the shock.
So the self-similar evolution of $F(Z)$ implies the constancy of $P_{c,2}$.
In this case $\tilde P_{c,2} \approx 0.31$ for $\tilde t \gsim 2$. 
>From that time forward the spatial distribution of
$P_c$ expands approximately linearly with time, as anticipated
in \S 2.4.
This demonstrates that the growth of a precursor 
and the shock structure proceed approximately in a self-similar way 
once the postshock CR pressure becomes constant.
In the lower panels of Fig. 4 we also show the evolution of 
$D(p)$ and $F(Z)$ for another model
with $M_0=50$, $T_0=10^4$K, $\tilde \kappa_{B}=0.01p/\sqrt{p^2+1}$
and $\theta=0.1$.  
For the this model  
$p_1\approx 2.0\times 10^{-3}$ and $p_{\rm max}\approx 10 \tilde t$. 
In this case, the CR pressure is mostly 
dominated by relativistic particles.

\begin{figure}
\begin{center}
\includegraphics*[height=40pc]{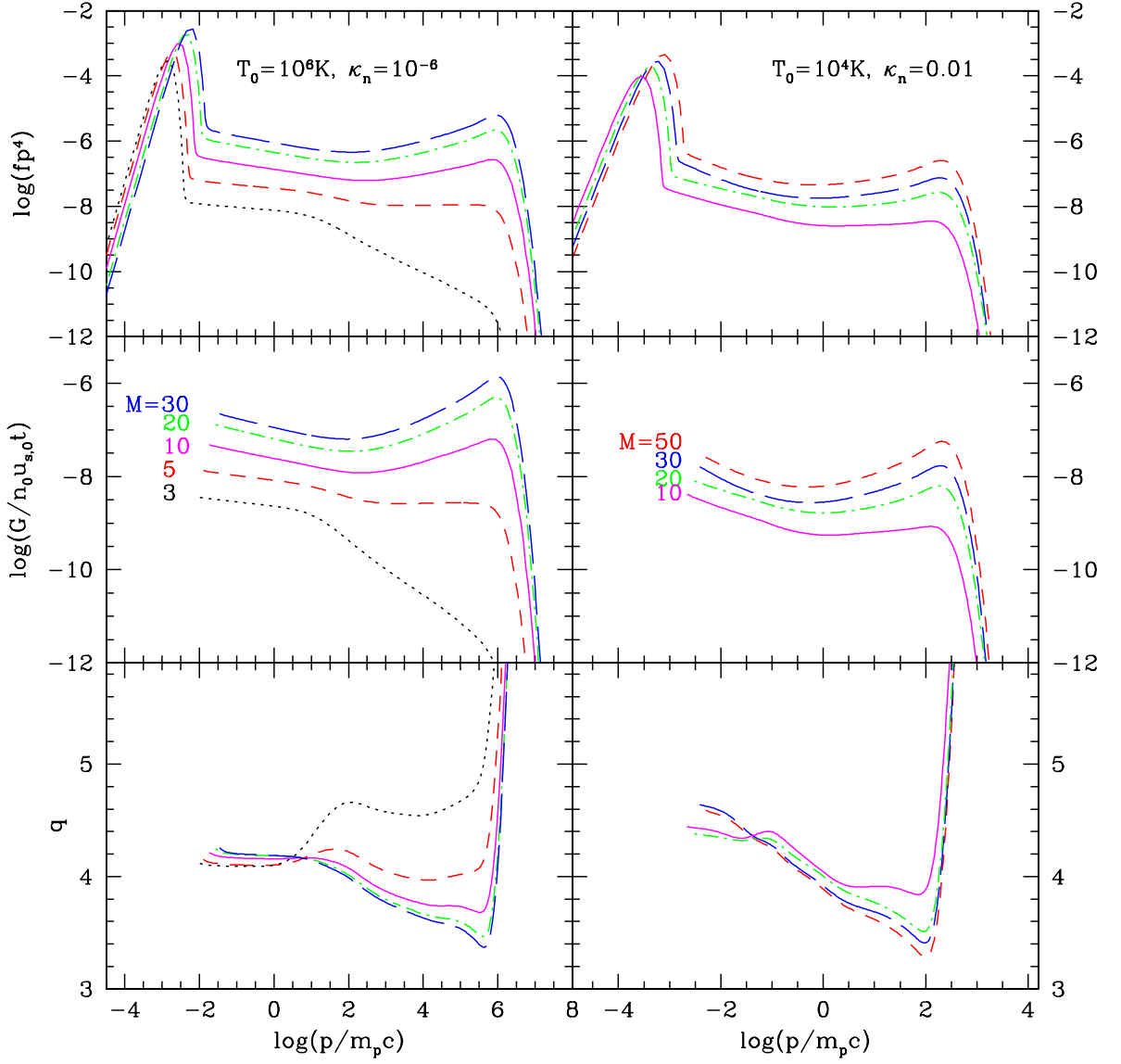}
\end{center}
\caption{
CR spectrum at the shock, $g(x_s,p)=f(x_s,p)p^4$, the volume
integrated distribution, $G=\int dx g(x,p)$, and $q=-d(\ln G)/d\ln p+4$ 
at $\tilde t=10$ are plotted for the models with $\theta=0.1$.
Left panels show models with $T_0=10^6$K, $\tilde \kappa_{BL}=10^{-6}p$,
and $3\le M_0 \le 30$,
while right panel show models with $T_0=10^4$K, 
$\tilde \kappa_{B}=0.01p/\sqrt{p^2+1}$, 
and $10\le M_0 \le 50$.
}
\label{fig5}
\end{figure}

Fig. \ref{fig5} compares the CR distributions at
$\tilde t = 10$ for two
sets of models spanning a range of sonic Mach numbers for different
diffusion model choices.
For the simulations represented on the left, the diffusion coefficient is 
Bohm-like, with
$\tilde \kappa_{BL}= 10^{-6}p$
and $T_0=10^6$K. The results on the right come from
a Bohm diffusion model with $\tilde \kappa_B = 0.01 p^2/\sqrt{p^2+1}$
and $T_0=10^4$ K.
For all models in this figure, $\theta=0.1$ and $\epsilon=0.2$.
The top panels show the CR distributions at the shock, $g(p,x_s) = p^4 f(p,x_s)$,
while the middle panels show the spatially integrated $G(p)$. 
The slopes of the integrated spectra, $q = - d (\ln G_p)/ d \ln p +4$,
are shown in the bottom panels.
For strong shocks with $M_0\ge 10$, both $p^4 f(p,x_s)$ and $G(p)$ 
exhibit concave curvature at high momentum
familiar from previous studies \eg \cite{berz99,maldru01,ama05}.
The hardened high momentum slopes reflect the fact that 
higher momentum CRs have longer mean free scattering paths,
so encounter an increased compression across the shock precursor
and a greater velocity jump. 

It is interesting to note in 
Fig. \ref{fig5} that the
spatially integrated spectra in the stronger shocks show more obvious 
hardening at high momenta
than do the spectra measured at the subshock. This is another consequence
of the fact that $l_d(p)$ increases with momentum, so that
the CR spectrum hardens considerably as one measures it
further upstream of the subshock. CRs escaping upstream from
such a shock would be dominated by the highest momentum particles
available.

\begin{figure}
\begin{center}
\includegraphics*[height=40pc]{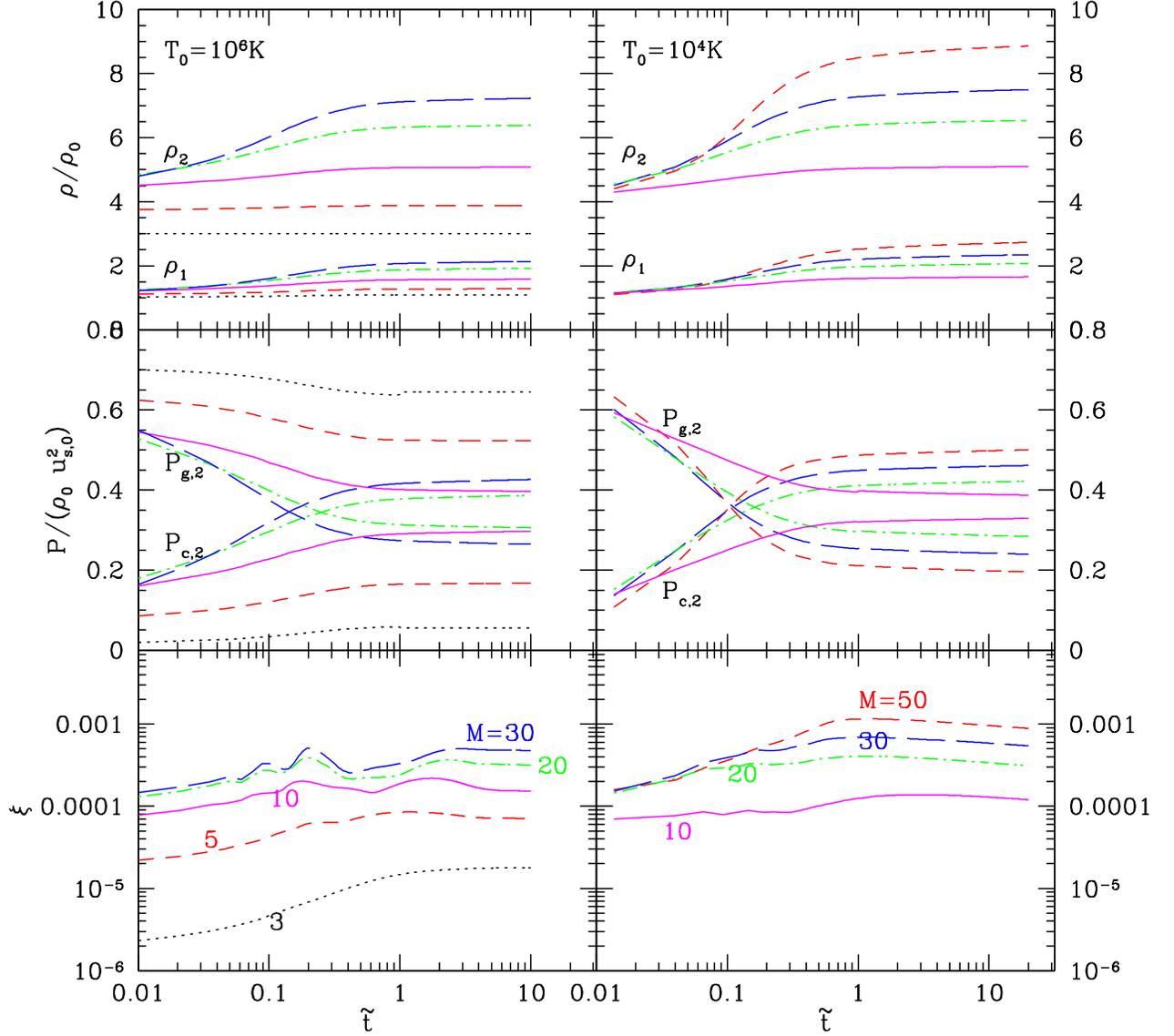}
\end{center}
\caption{
Precursor compression factor, $\sigma_p = \rho_1/\rho_0$, 
total compression factor, $\sigma_t = \rho_2/\rho_0$ ($\rho_0 = 1$),
postshock CR and gas pressure in units of the shock ram pressure,
and the injection efficiency $\xi$
are plotted as functions of time for the models shown in Fig. 5. 
}
\label{fig6}
\end{figure}

Fig. \ref{fig6} compares the evolution of shock properties for the
same set of models whose CR spectra are shown in Fig. \ref{fig5}.
The evolution of density increase through the precursor, 
$\sigma_p=\rho_1/\rho_0$,
and the total compression, $\sigma_t=\rho_2/\rho_0$,
are shown in each top panel. 
Compression through the subshock itself can be found through the ratio 
$\sigma_s = \rho_2/\rho_1$. 
We will discuss shock compression results in more detail below. 

The middle panels of Fig. \ref{fig6} show evolution of the postshock 
gas and CR pressures normalized to the ram pressure at the initial shock. 
After fast evolution at the start, these ratios approach constant values
for $\tilde t \gsim 1$. 
The bottom panels monitor the thermal leakage injection fraction, $\xi(t)$.
With the adopted value of $\epsilon_B=0.2$, the time asymptotic value of
this fraction is $\xi \sim 10^{-4} - 10^{-3} $ with the higher values
for stronger shocks. 
We note that the behavior of $\xi$ during the early phase ($\tilde t < 1$)
is controlled mainly by the injection suppression scheme used to prevent 
start up problems.
If the diffusion of particles with lowest momenta near $p_1$ is
better resolved with a finer grid spacing, one may observe 
some initial reduction of $\xi(t)$ as the subshock weakens in time (\eg
Fig. 6 of \cite{kjg02}).
The various plots show that postshock properties approach time-asymptotic
states that depend on the shock Mach number.
Generally, as is well known, the shock compression, the normalized
postshock CR pressure and the thermal leakage injection fraction
increase with shock Mach number. Nonlinear effects, illustrated here
by increased shock compression and diminished postshock gas pressure,
are relatively unimportant for Mach numbers less than 10 or so. At
high Mach number upstream gas temperature is relatively 
unimportant. In addition, noting that the simulations in the
left and right panels employed different models for the low
momentum diffusion coefficient, the similarity between the
analogous left and right plots illustrate the point made
earlier that asymptotic dynamical behaviors are insensitive to
details in the diffusion coefficient.

\begin{figure}
\begin{center}
\includegraphics*[height=40pc]{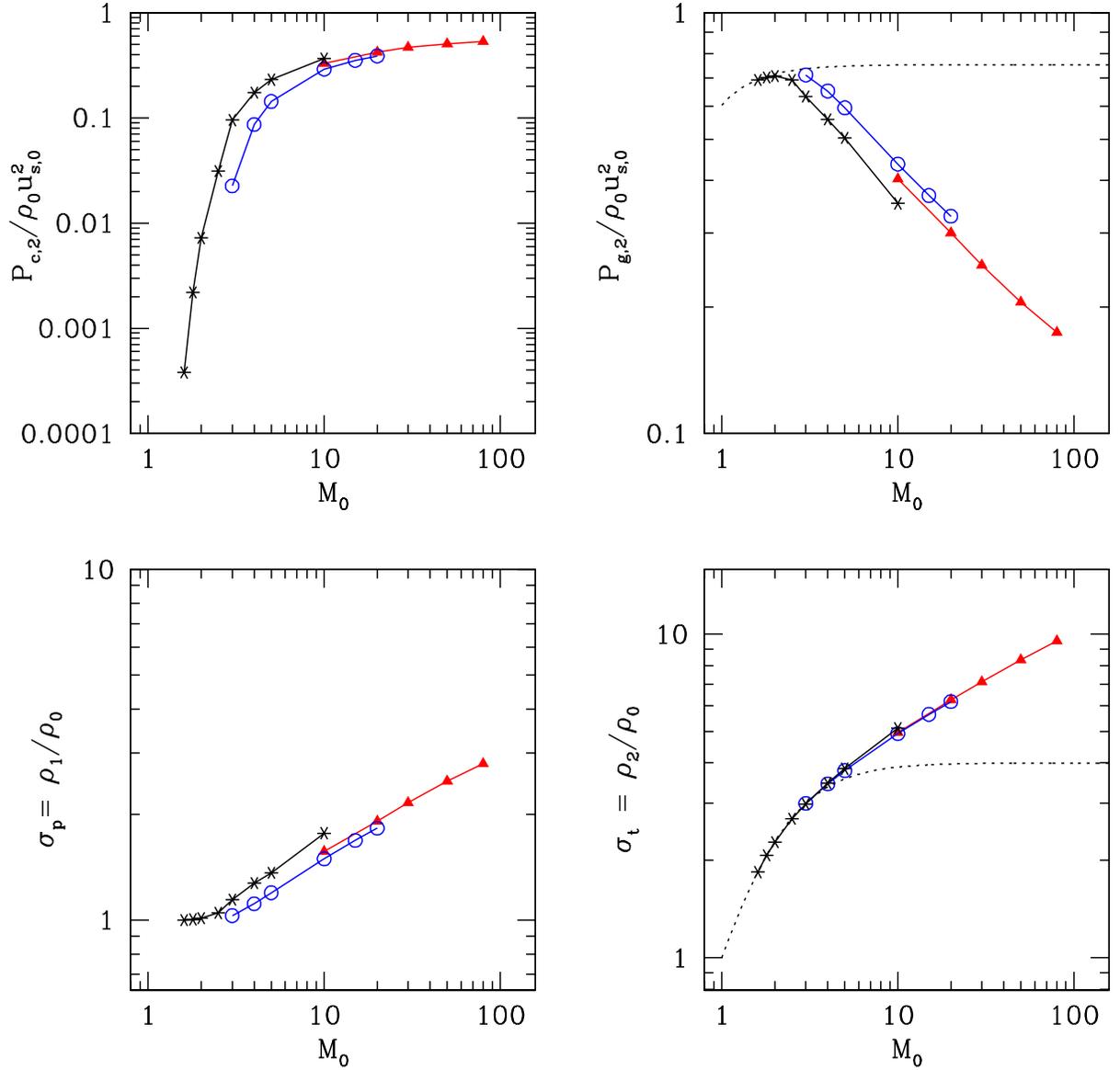}
\end{center}
\caption{
Time asymptotic values of
the postshock CR and gas pressure in units of the shock ram pressure,
precursor compression factor, $\sigma_p$, 
and total compression factor, $\sigma_t$, 
are plotted as functions of the shock Mach number. All models use
$\theta = 0.1$.
The dotted lines represent $P_{g,2}$ and $\sigma_t$
calculated from the Rankine-Hugoniot relation with $\gamma = 5/3$.
Symbols are used as follows: 
open circles for models with $T_0=10^6$K and $\epsilon_B=0.2$,
stars for models with $T_0=10^6$K and $\epsilon_B=0.25$,  
and filled triangles for models with $T_0=10^4$k and $\epsilon_B=0.2$.
}
\label{fig7}

\end{figure}

The Mach number dependences of several important shock dynamical
properties are illustrated in Fig. \ref{fig7}.
These include the
time-asymptotic values of $\sigma_t$, $\sigma_p$, and the postshock CR and
gas pressures, both normalized by the initial momentum flux,
$\rho_0 u^2_{s,0}$.
Dotted lines provide for comparison the $\gamma_{\rm g} = 5/3$
gasdynamic compression ratio and postshock pressure.
The strong shock limits of the gasdynamic compression ratio and
normalized postshock pressure are 4 and 0.75, respectively.
The two sets of models shown in Figs. 5-6 are shown,
along with an additional set of models with 
$T_0=10^6$K, $\kappa_{BL}=10^{-6}p$, $\theta=0.1$, and $\epsilon_B=0.25$.
The normalized postshock CR pressure increases with $M_0$, 
asymptoting towards $\tilde P_{c,2} \sim0.5$ at high Mach numbers.
These values are reduced $\sim$ 30\% from the analogous results
presented in \cite{kj05}, which is consistent with the comparison
for the Mach 10 shock shown in Fig. \ref{fig2}.
The normalized gas pressure simultaneously drops in response to
increases in $\tilde P_c$, 
as one would anticipate. It is notable that it actually falls 
well below 0.25, where one might naively expect it to asymptote
in order to maintain a constant total postshock pressure with respect
to $\rho_0 u_{s,0}^2$.
Instead it falls as
$\tilde P_{g,2} = P_{g,2}/(\rho_0 u_{s,0}^2) \sim 0.4 (M_0/10)^{-0.4}$,
without flattening at high Mach numbers.
As we shall see below, this trend is consistent with expected
evolution of gas in the precursor.
Thus {\it the total postshock pressure is less than that of a
gasdynamic shock of the same initial Mach number}, $M_0$, despite 
a softening of the equation of state coming from the CRs. 

For the $\epsilon_B = 0.2$ cases shown in Fig. \ref{fig7} 
the precursor, subshock, and total compression ratios can be approximated 
by $\sigma_p \sim 1.5(M_0/10)^{0.29}$,
$\sigma_s \sim 3.3 (M_0/10)^{0.04}$, and 
$\sigma_t = \sigma_p \sigma_s \sim 4.9(M_0/10)^{0.33}$, 
respectively, for $M_0 \gsim 5$.
For these models we also find the subshock Mach number ranges 
$3\lsim M_1 (= u_1/c_{s,1}) \lsim 4$ and depends on total Mach  
number roughly as $M_1 \propto M_0^{0.1}$. 
Since the Rankine-Hugoniot-derived gas shock compression in this Mach
number range scales as $\sigma_s \propto M_1^{0.35}$, so
$\sigma_s \propto (M_0^{0.1})^{0.35}$ is consistent with
the above relation.
Although weak, the variation of
the subshock strength with full shock sonic Mach number is important
to an understanding of the simulated flow behaviors seen in our shock
precursors.
We note that the subshock behavior seen here
in highly modified strong shocks, particularly that it
generally evolves towards Mach numbers close to three, is consistent 
with previous analytic and numerical results, 
although in some other studies the subshock strength is completely
independent of $M_0$ (\eg \cite{berz95,berz99,maldru01}).
The subshock strength in our simulations is determined by complex
nonlinear feedback involving the thermal leakage injection
process. As noted for the results in Fig. \ref{fig6}
the injection rate, $\xi$, generally increases with $M_0$.
This simultaneously increases $P_c$, providing extra
Alfv\'enic heating in the precursor, while cooling gas entering
the subshock, as energy is transferred to low energy CRs. It should
not be surprising that the balance of these feedback processes
is not entirely independent of total Mach number.
We comment in passing that the compression values shown  
in Fig. \ref{fig7} depend slightly on $\epsilon_B$ in the sense that
larger values of $\epsilon_B$ result in a little higher CR injection rate and
greater $P_c$.

Compression through the shock precursor has been discussed by several
previous authors (\eg \cite{berz95,berz97,mal97,kjg02}). We
outline the
essential physics to facilitate an understanding of our results. In a steady flow 
($\partial /\partial t = 0$) the modified entropy equation (\ref{scon})
can be integrated across the precursor with $\gamma_g = 5/3$ to give
\begin{equation}
\sigma_p = \frac{\rho_1}{\rho_0}=\left[\left(\frac{M_1}{M_0}\right)^2 + \frac{2}{3}
M_1^2~I\right]^{-3/8},
\label{compress}
\end{equation}
where $M_0 = u_{s,0}/c_{s,0}$ and
\begin{equation}
I = \frac{5}{3 u_0^3 \rho_0^{1/3}}\int\frac{|W|}{\rho^{2/3}}dx.
\label{diss}
\end{equation}
The same relation can be derived from a Lagrangian perspective
applying the second law of thermodynamics to
a parcel of gas flowing through the precursor, not necessarily in a steady
state.
For these simulations $|W| = |v_A \partial P_c/\partial x|$. Given
the previously mentioned result, $P_{c,2} = P_{c,1} \sim 0.5 \rho_0 u_s^2$ we can 
estimate that $I \approx v_A/u_s\approx 1.34 \sqrt{\theta}/M_0$. 
Eq. (\ref{compress}) is then
similar to an expression given in \cite{berz95}. When Alfv\'en wave
dissipation is small, so that $\theta << 5/(4M_0^2)$, Eq. 
(\ref{compress}) gives the result $\sigma_p \sim M_0^{3/4}$, appropriate
for adiabatic compression and consistent with behaviors found
in a number of previous analytic and numerical studies
(\eg \cite{berz95,berz97,mal97,kjg02}). 
For the simulations represented in Figs. \ref{fig3}-\ref{fig7} the
opposite limit actually applies, since $\theta = 0.1 > 5/(4M_0^2)$
whenever $M_0 > 3.5$. Then the strong Alfv\'en dissipation
limit of Eq. (\ref{compress})
predicts $\sigma_p \sim (M_0 \sqrt{\theta})^{3/8}/M_1^{3/4}$. Substituting our
observed relation between subshock Mach number and total Mach number, 
$M_1 \propto M_0^{0.1}$, we establish for fixed $\theta$
an expected behavior, $\sigma_p \propto M_0^{0.3}$, very close to 
what is observed. If instead of $\theta$ we had parameterized
the Alfv\'en wave influence in terms of the Alfv\'enic Mach number,
the analogous behavior of Eq. (\ref{compress}) would have been 
$\sigma_p \propto 1/(M_A^{3/8}M_1^{3/4})$.
Then for fixed Alfv\'enic Mach number the precursor compression
would vary only with the subshock Mach number, which, has only a
weak dependence on the full shock Mach number. That agrees with
results presented by \cite{berz95}, for example.

We mention that the precursor/subshock compression properties also explain
the observed inverse dependence of postshock gas pressure on
initial Mach number. In fact, in a steady flow it is easy to show that
\begin{equation}
\frac{P_{g,2}}{\rho_0 u_0^2} = \frac{2}{(\gamma_g + 1)}
\left(1 - \frac{(\gamma_g - 1)}{2 \gamma_g M_1^2}\right) 
\frac{1}{\sigma_p} \approx \frac{3}{4} \frac{1}{\sigma_p}.
\label{gasp}
\end{equation}
Thus we expect an inverse relation between precursor compression
and normalized postshock gas pressure, close to what is observed in our
simulations.

\begin{figure}
\begin{center}
\includegraphics*[height=40pc]{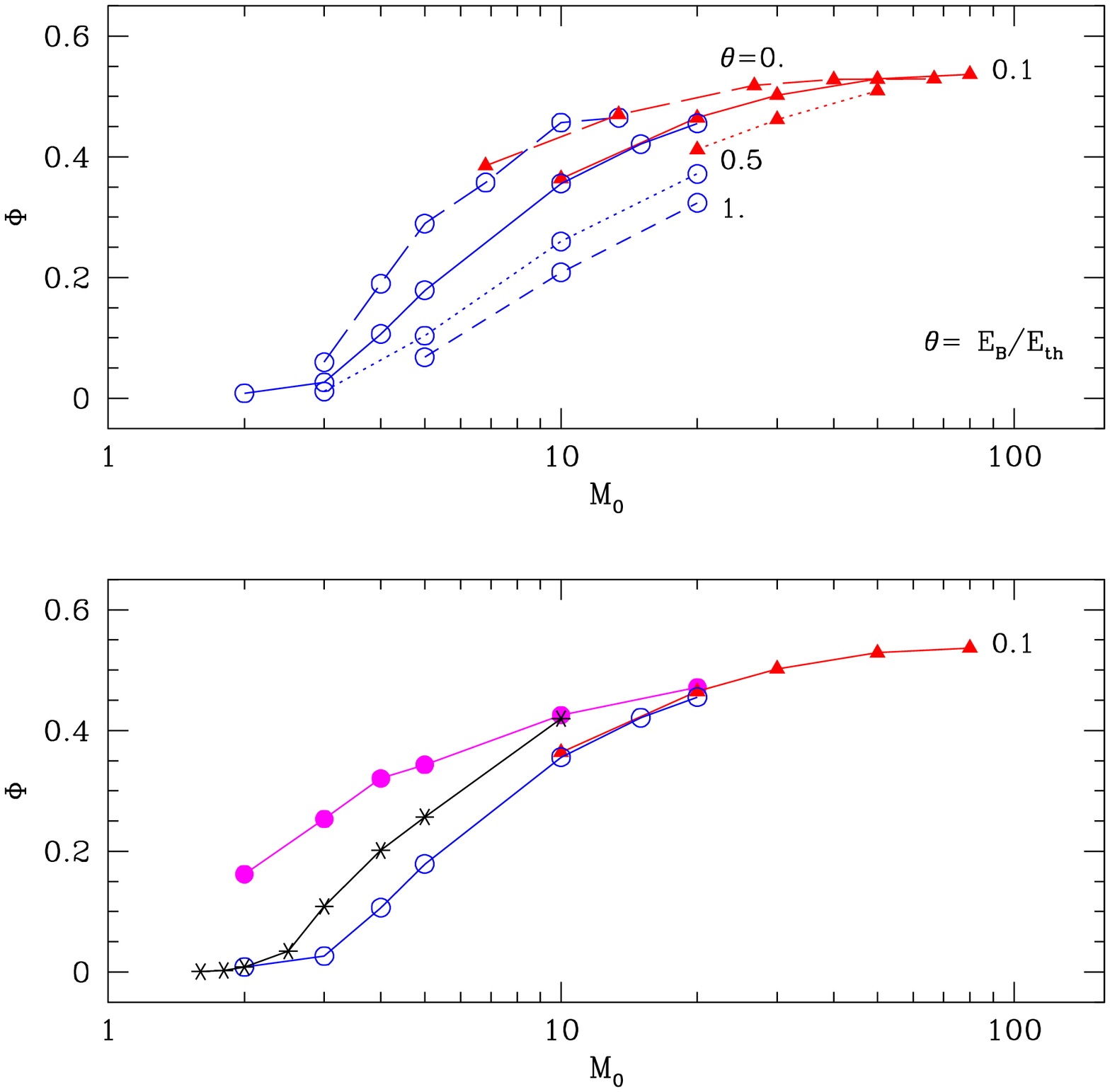}
\end{center}
\caption{
The CR energy ratio $\Phi$ is plotted as a function of the shock Mach number
for different models.
In the top panel, $\epsilon_B=0.2$ for all models,
open circles are used for models with $T_0=10^6$K,
while filled triangles are used for models with $T=10^4$K.
Long dashed lines are for $\theta=0.$,
solid lines for $\theta=0.1$,
dotted lines for $\theta=0.5$,
and short dashed lines for $\theta=1.$ cases.
In the bottom panel, 
models with open circles and filled triangles are the same 
as in the top panel and shown only to be compared with other cases.  
Filled circles connected with the solid line are for models with 
$T_0=10^6$K, $\epsilon_B=0.2$, and $\theta=0.1$ with a pre-existing
CR population with $P_{c,0}/P_{g,0}=0.25-0.5$ and $f(p)\propto p^{-5}$. 
Stars connected with the solid line:
$T_0=10^6$K, $\epsilon_B=0.25$, and $\theta=0.1$.  
}
\label{fig8}
\end{figure}

Finally, Fig. \ref{fig8} shows time-asymptotic values of the CR energy ratio, $\Phi(M_0)$, 
for models with different $\theta$ (top panel) and 
different $\epsilon_B$ and pre-existing CRs (bottom panel).
Models are shown with both $T_0=10^6$K and $T_0=10^4$K. 
This figure demonstrates that the CR acceleration depends 
on the specific model parameters considered here. 
For models with $\theta= 0.1$ and $T=10^6$ K,
the acceleration efficiency is reduced by up to $\sim50$ \% in comparison
to models without Alfv\'en wave drift and dissipation ($\theta = 0$).  
For larger Alfv\'en speeds with $\theta \sim 1$, the
acceleration efficiency is reduced even more significantly
as a consequence of strong preshock Alfv\'enic heating and
wave advection.  
On the other hand, for models with $T_0=10^4$K and $M_0>20$ 
the reduction factor is less than 15 \% for $\theta \sim 0.5$.
As shown in previous studies \cite{kjg02,kj02}, larger values of 
$\epsilon_B$ lead to higher thermal leakage injection and so more CR energy. 
Also the presence of preexisting CRs facilitates thermal leakage
injection, leading to more injected CR particles and higher acceleration
efficiency. 
Thus, an accurate estimate of the CR energy generated at quasi-parallel
shock requires detail knowledge of complex physical processes involved.
Fortunately, all theses dependences become gradually weaker 
at higher Mach numbers, and $\Phi$ tends to approach 0.5 for $M_0>30$.
It seems likely that this asymptotic efficiency would
also apply for sufficiently large $\theta$.
For low Mach number shocks, it is not yet possible to make simple, 
model-independent efficiency predictions.

\section{Conclusion and Summary}

While full thermalization takes place instantaneously at a simple,
discontinuous jump in an ideal gasdynamic shock, CR acceleration
and the corresponding modifications to the underlying flow
depend on suprathermal particles passing back and forth diffusively
across the shock structure via resonant scatterings with MHD waves.
So the time dependent evolution of CR modified shocks depend on complex
interactions between the particles, waves in the magnetic field, 
and underlying plasma flow.
These processes develop on the diffusion time scale and diffusion
length scale, which are expected to be increasing functions of
particle momentum, so CR acceleration and shock evolution rates 
slow over time.
Here we have carried out time-dependent DSA simulations to study the
evolution of plane CR modified shocks 
with quasi-parallel magnetic fields. 
Simple models for thermal leakage injection and Alfv\'en wave propagation
and dissipation are included. 
By adopting an underlying base grid that moves with the subshock,
our CRASH code with shock tracking and AMR techniques can achieve
good numerical convergence at a grid resolution
much coarser than that required in a fixed, Eulerian grid.
In the comoving grid, the shock remains at the same location,
so the compression rate is applied consistently to the CR distribution
at the subshock, resulting in much more accurate low energy CR acceleration.

We showed that the time dependent evolution of CR modified shocks becomes 
approximately self-similar, because the postshock variables including the
CR pressure approach time-asymptotic values 
while the shock structure broadens linearly with time
as $l_{\rm shock}\sim u_s t/8$ and the maximum CR momentum increases
approximately as $p_{\rm max} \approx l_{\rm shock} u_s / \kappa_n$.
This behavior develops because the thermal leakage injection rate is controlled mainly 
by the shock Mach number and
because the ratio $t_{acc}/t_d$ is a function of shock Mach number.
The self-similar behavior is independent of the assumed diffusion
coefficient. 
When the drift and dissipation of Alfv\'en waves in the upstream region
are included, 
the CR acceleration is less efficient for a given Mach number, as expected
from previous studies \cite{jon93,berz97}. However, at
sufficiently high Mach numbers it appears, even with the
inclusion of the Alfv\'en wave terms, that the efficiency probably
asymptotes to approximately 50\% as measured by both the
fraction of energy flux transferred to CRs and the fraction of
momentum flux that appears as postshock CR pressure.

We summarize here the main properties of our improved simulations including
Alfv\'en wave effects:
 
1. The postshock pressures, $P_c$ and $P_g$,
approach time-asymptotic values quickly. 
At sufficiently strong shocks the postshock CR pressure is about 50 \% of the shock ram pressure,
while the normalized gas pressure drops from its gasdynamic value of 75 \%.
The normalized postshock gas pressure scales approximately inversely
with the compression through the shock. The resulting total
pressure in a strong CR modified shock will be reduced from
that in an ordinary gas shock of the same initial Mach number.
Typically a fraction of $\xi= 10^{-4} - 10^{-3}$ of the incoming thermal
particles become CRs through thermal leakage.

2. A significant shock precursor develops in response 
to nonlinear feedback from the CR pressure, so that the subshock weakens 
to $M_{\rm subshock} \sim 3$ for the initial shock Mach number
$ M_0 \ge 5$. The subshock strength increases very slowly with
total Mach number in a balance between energy extracted by CRs
and heating of the precursor.

3. In our present simulations heating in the precursor is dominated
in strong shocks
by Alfv\'en wave dissipation. The resulting asymptotic
density compression factor through the precursor can 
approximated by $\sigma_p \sim 1.5(M_0/10)^{0.29}$ for $M_0 \gsim 3$,
while the compression over the total transition can be approximated by
$\sigma_t \sim 4.9(M_0/10)^{0.33}$ for $M_0 \gsim 5$.
This behavior is consistent with simple theoretical expectations.
Of course, both $\sigma_s$ and $\sigma_t$ take gasdynamic values 
for smaller $M_0$.

4. Both the CR momentum distribution at the shock, $g(x_s,p)=f(x_s, p)p^4$, 
and the integrated distribution, $G(p)= \int {\rm dx} g(x,p)$, clearly
exhibit characteristic concave curvature, 
reflecting the nonlinear velocity structure in the precursor.
At non-relativistic momenta $G(p) \propto p^{-\alpha}$ with
$ 4.5<\alpha<4.2$, reflecting the strength of weaker subshocks,
while it flattens toward the highest momenta. Despite strong nonlinearity
in the shock structures, the maximum CR momentum still scales
reasonably close to predictions from test-particle theory. This behavior
allows
the shock precursor to evolve in an approximately self-similar
manner.

5.  The CR energy ratio $\Phi(Ms)$ increases with $M_0$ for modest Mach numbers
and depends on $\epsilon_B$, and the ratio $\theta=E_B/E_{th}$.
At sufficiently high Mach number $\Phi$ asymptotes towards $ \sim 0.5$.
Since acceleration efficiency is reduced by Alfv\'en wave effects,
higher Mach numbers are required to approach this limit as $\theta$
increases.
For larger $\epsilon_B$ in the thermal injection model, the magnetic 
turbulence is weaker, so that
particles can stream upstream more easily. 
CR injection is then more efficient. The importance of this
effect is significant only in relatively weak shocks.
The presence of pre-existing CRs is equivalent to a higher injection rate.
In weak shocks this translates into an apparently enhanced acceleration
efficiency as measured by the fraction of energy incorporated
into CRs. This influence is small in strong shocks, however.

\section*{Acknowledgments}
The authors would like to thank D. Ryu for suggesting the idea of
a comoving grid for CRASH code.
HK was supported by the Korea Research Foundation Grant
funded by Korea Government (MOEHRD, Basic Research Promotion Fund)
(R04-2004-000-100590)
and by KOSEF through the Astrophysical Research Center
for the Structure and Evolution of Cosmos (ARCSEC).
TWJ is supported at the University of Minnesota
by NASA grant NNG05GF57G, NSF grant Ast-0607674
and by the Minnesota Supercomputing Institute.




\begin{thebibliography}{00}




\bibitem{ach82}
Achterberg, A. Astron. \& Astrophys. 98 (1982) 195

\bibitem{achbl86}
Achterberg, A. \& Blandford, R. D. M.N.R.A.S. 218 (1986) 551

\bibitem{aharon04}
Aharononian, F. A. \etal Nature 432 (2004) 75

\bibitem{ama05}
Amato, E. \& Blasi, P. M.N.R.A.S. 364 (2005) L76

\bibitem{beck01}
Beck, R. Space Sci. Rev. 99 (2001), 243

\bibitem{bell78}
Bell, A. R. M.N.R.A.S. 182 (1978) 147

\bibitem{berz95}
Berezhko E.G., Ksenofontov, L. \& Yelshin, V. Nuc. Phys. B 39A (1995) 171

\bibitem{berz97}
Berezhko, E.G. \& V\"olk, H.J. Astroparticle Physics 7  (1997) 183

\bibitem{berz99}
Berezhko E. G. \& Ellison, D. C. Astrophys. J. 526 (1999) 385

\bibitem{blaeic87} 
Blandford, R.~D., and Eichler, D. Phys. Rept. 154 (1987) 1

\bibitem{blasi06}
Blasi, P. Amato, E. \& Caprioli, D. M.N.R.A.S. 375 (2007) 1471

\bibitem{car02}
Carilli C. L., \& Taylor G. B. ARAA 40 (2002) 319

\bibitem{dru83} 
Drury, L.~O'C. Rept. Prog. Phys. 46 (1983) 973

\bibitem{dvb95}
Drury, L. O'C., V\"olk, H. J. \& Berezhko, E. G. Astron. \& Astrophys. 299 (1995) 299

\bibitem{ddk96}
Drury, L. O'C., Duffy, P. \& Kirk, J. G. Astron. \& Astrophys. 309 (1996) 1002

\bibitem{ebj95} 
Ellison, D. C., Baring, M. G. \& Jones, F. C. 1995, Astrophys. J. 453 (1995) 873

\bibitem{elber99}
Ellison, D. C. \& Berezhko, E. G. 26th ICRC 4 (1999) 446

\bibitem{ellison05} 
Ellison, D. C., Decourchelle, A., Ballet, J. Astron. \& Astrophys. 429 (2005), 569

\bibitem{ell85}
Ellison, D. C. \& Eichler, D. Phys. Rev. Lett. 55 (1985) 2735

\bibitem{Fabian94} 
Fabian, A. C. ARAA, 32, (1994) 277

\bibitem{gies00}
Gieseler, U. D. J., Jones, T. W. \& Kang, H. Astron. \& Astrophys. 364 (2000) 911

\bibitem{jon93}
Jones, T.~W. Astrophys. J. 413 (1993) 619

\bibitem{kj91}
Kang, H. \& Jones, T. W., M.N.R.A.S. 249 (1991) 439

\bibitem{kjr92} 
Kang, H., Jones, T.~W., and Ryu, D. Astrophys. J. 385 (1992) 193

\bibitem{kjls01} 
Kang, H., Jones, T. W., LeVeque, R. J., Shyue, K. M. Astrophys. J. 550 
(2001) 737

\bibitem{kjg02} 
Kang, H., Jones, T. W., \& Gieseler, U.D.J. Astrophys. J. 579 (2002) 337

\bibitem{kj02}
Kang, H. \& Jones, T. W., Journal of Korean Astronomical Society, 35 (2002), 159

\bibitem{kang03}
Kang, H., Journal of Korean Astronomical Society, 36 (2003), 1
 
\bibitem{kj05} 
Kang, H., \& Jones, T. W. Astrophys. J. 620 (2005) 44 

\bibitem{kj06} 
Kang, H., \& Jones, T. W. Astropart. Phys. 25 (2006) 246 

\bibitem{lc83}
Lagage, P. O. \& Cesarsky, C. J. Astron. \& Astrophys. 125 (1983) 249

\bibitem{lucek00}
Lucek, S. G. \& Bell, A. R. M.N.R.A.S. 314 (2000) 65

\bibitem{mal97}
Malkov, M. A. Astrophys. J. 485 (1997) 638

\bibitem{mal98} 
Malkov M.A. Phys. Rev. E,  58 (1998) 4911

\bibitem{maldi06}
Malkov, M. A. \& Diamond, P. H. Astrophys. J. 642 (2006) 244

\bibitem{maldru01} 
Malkov, M.A., \& Drury, L.O'C. Rep. Progr. Phys. 64 (2001) 429

\bibitem{malvol98} 
Malkov, M.A., and V\"olk, H.J. Adv. Space Res. 21 (1998) 551

\bibitem{McKee77} 
McKee, C. F., and Ostriker, J. P.  ApJ, 218 (1977) 148

\bibitem{pfromer06}
Pfrommer, C., Springel, V., Ensslin, T. A., \& Jubelgas, M.,
Mon. Not.R. Astron. Soc., 367 (2006) 113

\bibitem{ptus03}
Ptuskin, V. S. \& Zirakashvili, V. N. Astron. \& Astrophys. 403 (2003) 1

\bibitem{rott97}
R\"ottgering, H.J.A., Wieringa, M.H., Hunstead, R. W. \& Ekers, R.D M.N.R.A.S. 290 (1997) 577

\bibitem{skill75}
Skilling, J., Mon. Not.R. Astron. Soc., 223 (1975) 353

\bibitem{Spitzer90} 
Spitzer, L., ARAA, 28 (1990) 71

\bibitem{vlad06}
Vladimirov, A., Ellison, D. C, \& Bykov, A. Astrophys. J. 652 (2006) 1246

\end{thebibliography}
\end{document}